# Retrieval of missing small-angle scattering data in gas-phase diffraction experiments

Yanwei Xiong*, Nikhil Kumar Pachisia, Martin Centurion†

Department of Physics and Astronomy, University of Nebraska-Lincoln, Lincoln, Nebraska, 68588, USA

**ABSTRACT**. We report an iterative algorithm to retrieve accurate real space information from gas phase ultrafast diffraction measurements with missing data at low momentum transfer. The algorithm operates in a manner similar to phase retrieval algorithms which transform signals back and forth between the signal domain and the Fourier transform domain and apply constraints to retrieve the missing phase. Our algorithm retrieves missing data that is not accessible experimentally by applying a real-space constraint that requires only an approximate a-priori knowledge of the shortest and longest internuclear distances in the molecule. We demonstrate successful retrieval of the missing data in simulated data and in experimentally measured electron diffraction signals of photo-dissociated iodobenzene molecules.

## I. INTRODUCTION.

Gas phase ultrafast electron diffraction (GUED) and ultrafast X-ray diffraction (UXRD) have demonstrated the capability to determine the structures of isolated molecules and to capture their nuclear motions during chemical reactions with femtosecond and sub-angstrom resolutions [1, 2]. In GUED and UXRD experiments, a pump laser is used to excite the molecules and a probe pulse, either an electron pulse with a kinetic energy of tens of keV [3-6] or MeV [7-10] or an X-ray pulse with an energy around 10 keV [11-14], is used to interrogate the evolving structure of the photoexcited molecules. The small scattering probability in gas targets leads to most electrons or photons passing straight through the sample without interactions. A beam stop or hole is required to prevent the high intensity transmitted beam from striking the detector to prevent saturation of the image and damage of the detector. This results in the loss of the signal at low scattering angles. Additionally, for GUED experiments from ions [15-17] or molecules excited to high-lying states [18], an additional signal appears in the low momentum transfer region which is due to changes in the electronic density. In these cases, the signal at low scattering angles cannot be used to generate the real-space pair distribution function (PDF) even if it is captured experimentally, because it contains mixed signals from electronic and nuclear motions.

The maximum momentum transfers accessible for GUED and UXRD experiments are limited by the detector size and/or by the signal levels which decrease rapidly with increasing scattering angle, or in the case of X-rays also by the wavelength. Therefore, the experimentally available range of momentum transfer is limited by a minimum value $s_{min}$ and maximum value $s_{max}$. The value of $s_{max}$ determines the spatial resolution of the measurement but does not introduce artifacts in the retrieval if treated appropriately.

Different methods have been developed to address the issue of the missing low-angle signal. In many cases the real-space analysis is avoided and the signal in momentum space is used to compare with theory. Agreement between the experimental and calculated signals is used to validate the theoretical structures produced by trajectory simulations [13, 19-23]. This, however, prevents the direct interpretation of the experimental signals in real space, and limits the possible structures retrieved to those that can be simulated. The second method is to fill the missing signal from 0 Å$^{-1}$ to $s_{min}$ by smooth interpolation to zero at the origin, and then interpret the data in real space by calculating the PDF [24, 25]. While this method avoids biasing structure retrieval with theory, it can introduce artifacts in the PDF. The third method is to fill the missing region with the signal produced by simulations [3, 26-30]. This has the disadvantage that it requires expensive calculations and may bias the comparison of experimental and theoretical PDFs. More recently, a model-free inversion technique that produces a super-resolved PDF from data that is limited in momentum transfer has been developed [31] and applied to experimental data [32]. However, this method requires a sparse signal, and thus its applicability is limited. Another alternative is the use of a genetic algorithm to retrieve a structure directly from the experimental data, without generating a PDF [18, 33-35]. This method requires *a-priori* knowledge to limit the number of possible structures and is challenging to apply to reactions with multiple channels.

In this work, we introduce an iterative algorithm that retrieves the diffraction signal in the missing region from 0 Å$^{-1}$ to $s_{min}$ and requires only an approximate *a-priori* knowledge of the minimum and maximum interatomic distances in the molecular structure, which is often readily available. The method does not require that a single structure is present in the signal and can be applied to experiments with multiple reaction channels. The iterative algorithm is adapted from the ideas of existing phase retrieval algorithms, like the Gerchberg-Saxton algorithm, the error-reduction algorithm and the hybrid input-output algorithm [36-41]. These algorithms iteratively transform the data back and forth between the object domain and the Fourier domain through Fourier transform and inversion, applying known constraints in each domain while allowing the phase to be adjusted until a solution is found [36, 42]. These ideas have also been used to reconstruct X-ray images of objects with a size of tens of nanometers from blurred diffraction images [41, 43]. Our application is different in that we do not retrieve the phase of the scattering signal but the missing data which enables us to calculate the real-space PDF. The algorithm iterates between the momentum transfer and real spaces and applies a support constraint (limited size) in real space, which is sufficient to accurately retrieve the missing data.

We demonstrate the accuracy of the retrieval algorithm with simulated and experimental electron scattering signals of ground state and dissociated iodobenzene, including diffraction signals of unexcited and dissociated molecules. We have selected this molecule because it is comparable in size to many of the samples used in GUED and UXRD experiments. In addition to leading to accurate retrieval of the PDFs, this algorithm could also be used to separate the electronic and nuclear contributions to the GUED signals.

## II. THEORY

In this section, we first review the electron scattering theory for gas phase molecules that is relevant for this work and then describe the iterative algorithm.

### A. Electron scattering theory

The elastic scattering from a neutral molecule can be well approximated using the independent atom model (IAM), in which the bonding effects are neglected and the potential of each atom in the molecule is assumed to be spherically symmetric [15, 18, 44]. Here we focus on restoring the missing signal in an electron diffraction signal produced by a sample of randomly oriented molecules. The total scattering intensity $I_{total}(s)$ consists of the atomic scattering $I_A(s)$ and the molecular scattering intensity $I_M(s)$ [45, 46].

$$I_{total}(s) = I_A(s) + I_M(s), \quad (1)$$

$$I_A(s) = \sum_{j=1}^{N} \left| f_j(s) \right|^2, \quad (2)$$

$$I_M(s) = \sum_{j=1}^{N} \sum_{k \neq j}^{N} f_j^*(s) f_k(s) \frac{\sin(sr_{\mathrm{jk}})}{sr_{\mathrm{jk}}}, \quad (3)$$

where $N$ is the number of atoms that constitute the molecule, $s$ is the momentum transfer with magnitude $s = \frac{4\pi}{\lambda}\sin\left(\frac{\theta}{2}\right)$, in which $\lambda$ is the wavelength and $\theta$ is the scattered angle of electrons, $f_j(s)$ is the atomic scattering amplitude of the $j^{\text{th}}$ atom [47, 48], and $r_{ij}$ is the interatomic distance between the $\mathrm{i}^{\text{th}}$ and the $\mathrm{j}^{\text{th}}$ atoms. The atomic scattering $I_A(s)$ contains no structural information, whereas the molecular scattering $I_M(s)$ encodes information about the interatomic distances. Since the amplitude of $I_M(s)$ is approximately proportional to $s^{-5}$, the modified scattering intensity $sM(s)$ is introduced to compensate the rapid decrease in amplitude [28, 49]:

$$sM(s) = \frac{sI_M(s)}{I_{\mathrm{A}}(s)}. \quad (4)$$

The interatomic distances $r_{jk}$ can be obtained (within the spatial resolution of the measurement) by applying a Fourier sine transform (FST) to the $sM(s)$ to produce the PDF [28, 49], given by

$$\mathrm{PDF}(r) = \int_0^{s_{max}} sM(s)\sin(sr)e^{-ds^2}ds. \quad (5)$$

Where $e^{-ds^2}$ is a damping function used to avoid artifacts due to the discontinuity of the signal at $s_{max}$.

GUED measurements probe changes of the molecular structure using the diffraction difference intensity

$$\Delta I_M(s,t) = I_{total}(s,t) - I_0(s), \quad (6)$$

where the variable $t$ denotes the time delay with respect to the laser excitation. $I_0(s)$ is a reference signal corresponding to the ground state (unexcited) molecules [27, 50]. Correspondingly, the $\Delta sM(s)$ can be defined as

$$\Delta sM(s,t) = \frac{s\Delta I_{\mathrm{M}}(s,t)}{I_{\mathrm{A}}(s)}. \quad (7)$$

The $\Delta\mathrm{PDF}(r)$ can be calculated by replacing $sM(s)$ in eqn (5) with $\Delta sM(s,t)$.

### B. Retrieval method

In this section, we describe the iterative algorithm to restore the missing data. First, we establish the FST pairs by defining an odd function that corresponds to the *sM*. Second, we mathematically model the typical experimental electron diffraction signal with a limited momentum transfer from $s_{min}$ to $s_{max}$ and the artifact that appears in real space due to the missing data. Third, we apply a band-pass filter as a support constraint in real space that iteratively reduces the amplitude of the artifact.

Based on eqn (4) and eqn (5), we define an odd function $\mathcal{M}(s)$ for the static diffraction signal by extending the variable $s$ from $0 \leq s \leq s_{max}$ to

$-s_{max} \leq s \leq s_{max}$. The odd function $\mathcal{M}(s)$ is given by

$$\mathcal{M}(s) = \begin{cases} \sum_{j=1}^{N} \sum_{k \neq j}^{N} c_{jk}(s) \frac{\sin(s r_{jk})}{r_{jk}} & \text{for } |s| \leq s_{max} \\ 0 & \text{otherwise} \end{cases} \quad (8a)$$

where $c_{jk}(s) = f_j^*(s) f_k(s) / I_A(s)$. Note that the negative *s* is only mathematically meaningful. For the diffraction difference signal $\Delta sM$ used in time-resolved experiments, $\mathcal{M}(s)$ is given by

$$\mathcal{M}(s) = \begin{cases} \sum_{j=1}^{N} \sum_{k \neq j}^{N} c_{jk}(s) [\frac{\sin(s \tilde{r}_{jk})}{\tilde{r}_{jk}} - \frac{\sin(s r_{jk})}{r_{jk}}] & \text{for } |s| \leq s_{max} \\ 0 & \text{otherwise} \end{cases} \quad (8b)$$

where $\tilde{r}_{jk}$ are the newly produced interatomic distances after laser excitation, and $r_{jk}$ are the interatomic distances before laser excitation. The function $\mathcal{M}(s)e^{-ds^2}$ is an odd function, and the FST and Fourier sine inverse transform (FST$^{-1}$) are

$$\mathcal{P}(r) = \int_0^{+\infty} \mathcal{M}(s) e^{-ds^2} \sin(sr) ds, \quad (9)$$

$$\mathcal{M}(s) e^{-ds^2} = \frac{2}{\pi} \int_0^{+\infty} \mathcal{P}(r) \sin(sr) dr. \quad (10)$$

The $sM(s)$ given by eqn (4) or $\Delta sM$ by eqn (7) is identical to the function $\mathcal{M}(s)$ for $s \geq 0$. Therefore $\mathcal{P}(r)$ in eqn (9) is equal to the PDF$(r)$ or $\Delta$PDF$(r)$. Suppose the missing signal in the range 0 Å$^{-1}$ to $s_{min}$ is filled by a first-guess function $\mathcal{G}(s)$, which could be for example zeros, a linear fit, quadratic fit, *etc*. Similarly to eqn (8a) and eqn (8b), an odd function $\mathcal{M}_e(s)$ that corresponds to the measured diffraction signal can be defined as

$$\mathcal{M}_e(s) = \begin{cases} \mathcal{G}(s) & \text{for } |s| < s_{min} \\ \mathcal{M}(s) & \text{for } s_{min} \leq |s| \leq s_{max} \\ 0 & \text{otherwise} \end{cases} \quad (11)$$

The FST of $\mathcal{M}_e(s)e^{-ds^2}$ is $\mathcal{P}_e(r)$, which corresponds to the direct transform of the experimental signal with incorrect information in the region $|s| < s_{min}$. The $\mathcal{P}_e(r)$ can be written as

$$\mathcal{P}_e(r) = \mathcal{P}(r) + \mathcal{A}(r), \quad (12)$$

Where $\mathcal{P}(r)$ is the true signal that we are seeking and $\mathcal{A}(r)$ is an artifact in real space due to the difference between $\mathcal{M}_e(s)$ and the true signal $\mathcal{M}(s)$. $\mathcal{A}(r)$ can be expressed as

$$\mathcal{A}(r) = \int_0^{s_{min}} [\mathcal{G}(s) - \mathcal{M}(s)] \, e^{-ds^2} \sin(sr) ds. \quad (13)$$

Also, we have the following relation

$$[\mathcal{M}_e(s) - \mathcal{M}(s)] e^{-ds^2} = \frac{2}{\pi} \int_0^{+\infty} \mathcal{A}(r) \sin(sr) dr. \quad (14)$$

Our goal is to iteratively reduce the amplitude of $\mathcal{A}$(r) to retrieve $\mathcal{M}(s)$. The two conditions for a successful retrieval are: (1) The distribution of $\mathcal{A}$(r), which corresponds to the missing signal from 0 Å$^{-1}$ to $s_{min}$, is broader than that of $\mathcal{P}$(r). (2) Prior knowledge of the minimum and maximum interatomic distances in the molecule and the reaction products, which in some cases can be estimated from the data itself. It is difficult to establish general conditions on the required s-range for the retrieval because the constraint is applied in real space and the distribution of distances will depend on the structure of each molecule and the nature of the reaction.

Figure 1 shows a block diagram of the algorithm with iteration number denoted by *n*. The steps are as follows.
(1) At the start, for *n*=1, we set $\widetilde{\mathcal{M}}_1(s) = \mathcal{M}_e(s)$.
(2) A FST of $\widetilde{\mathcal{M}}_n(s)e^{-ds^2}$ generates $\mathcal{P}_n(r)$.
(3) The real-space support constraint, based on prior knowledge, is implemented by applying a band-pass filter $\mathcal{H}$(r) that limits the minimum and maximum distances to produce $\tilde{\mathcal{P}}_n(r)$. The key idea here is that $\mathcal{P}(r)$ is constrained within filter, while $\mathcal{A}(r)$ will extend beyond the filter so the amplitude of the artifact will be reduced with each iteration.
(4) A FST$^{-1}$ is applied to $\tilde{\mathcal{P}}_n(r)$ to generate $\mathcal{M}_{n+1}(s)e^{-ds^2}$.
(5) $\mathcal{M}_{n+1}(s < s_{min})$ and $\mathcal{M}_e(s_{min} \leq s \leq s_{max})$ are stitched together to produce $\widetilde{\mathcal{M}}_{n+1}(s)$, in which the signal from 0 to $s_{min}$ is closer to the true signal compared to the previous iteration. To avoid a discontinuity in $\widetilde{\mathcal{M}}_{n+1}(s)$, the signal $\mathcal{M}_{n+1}(s < s_{min})$ is multiplied by a rescaling factor given by $\int_{s_{min}}^{s_{min}+\varepsilon} \mathcal{M}_e(s) ds / \int_{s_{min}-\varepsilon}^{s_{min}} \mathcal{M}_{n+1}(s) ds$, where $\varepsilon$ is a small value, such as 0.01 Å$^{-1}$.
(6). Repeat step (2) by replacing $\widetilde{\mathcal{M}}_n(s)$ with $\widetilde{\mathcal{M}}_{n+1}(s)$ to generate $\mathcal{P}_{n+1}(r)$ and $\tilde{\mathcal{P}}_{n+1}(r)$, followed by steps (3)-(5).

The retrieval error is defined as the sum of the square of the difference between $\mathcal{M}_n$ and $\mathcal{M}_e$:

$$\mathcal{S}_n = \frac{1}{s_{max} - s_{min}} \int_{s_{min}}^{s_{max}} [\mathcal{M}_n(s) - \mathcal{M}_e(s)]^2 ds. \quad (15)$$

$\widetilde{\mathcal{M}}_n(s)$ approaches $\mathcal{M}(s)$ as $\mathcal{A}(r)$ approaches zero, according to eqn (14). The algorithm is stopped when $\mathcal{S}_n$ decreases to a small number and reaches a plateau.

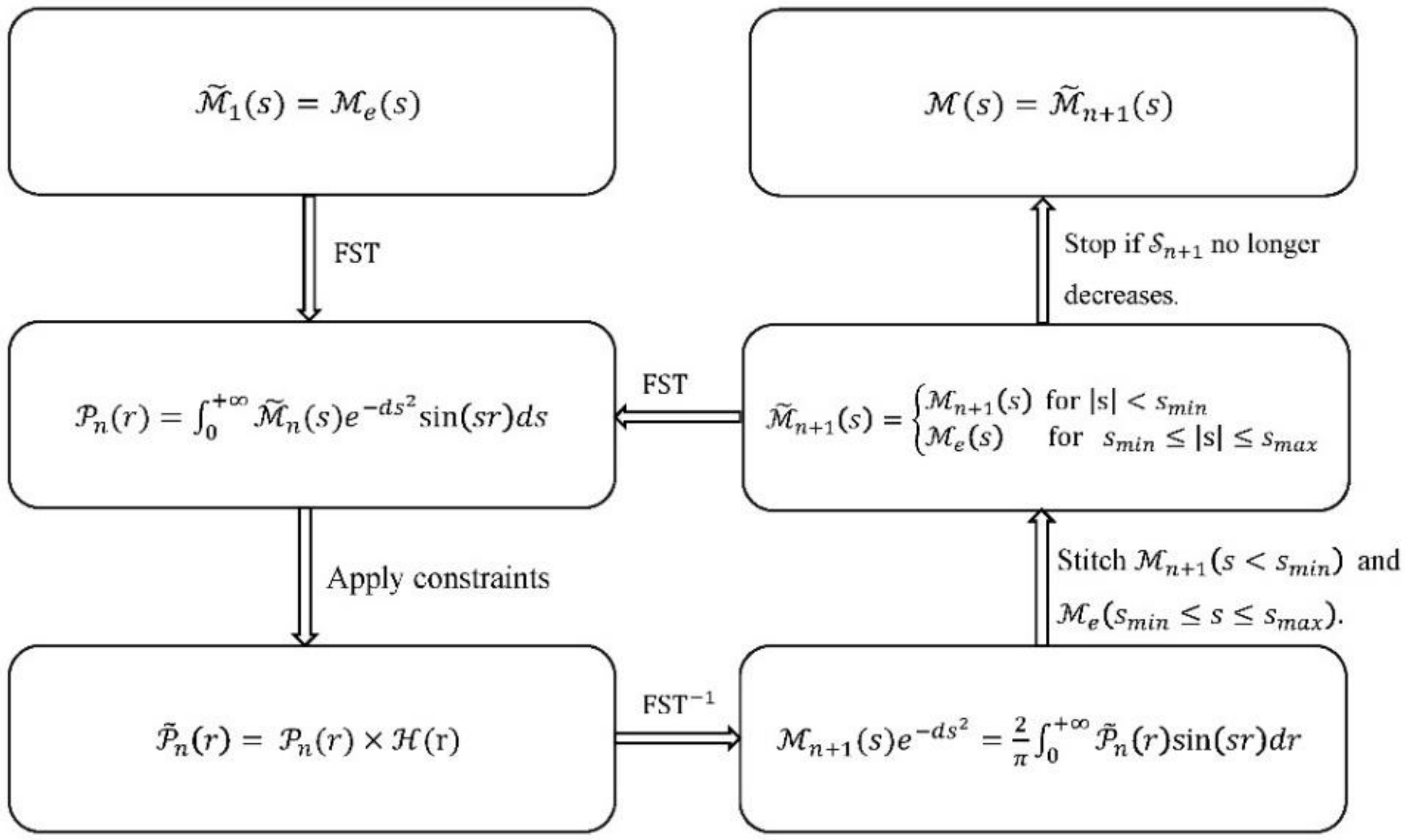

Figure 1. Block diagram of the iterative algorithm. The iteration number is denoted by $n$. FST stands for Fourier sine transform, and $\mathrm{FST}^{-1}$ stands for the Fourier sine inverse transform.

## III. TEST WITH SIMULATED DATA

In this section, we test the iterative algorithm using a calculated diffraction signal. The kinetic energy of the electrons used in the calculation is 90 keV, and the scattering amplitude of the atoms are tabulated in [51]. We use eqns (3) and (4) to calculate the $sM(s)$ and select the $s$-range 1.6 Å$^{-1}$ $\leq s \leq$ 10 Å$^{-1}$ of the data as the available signal, matching the typical range in GUED measurements [9, 52]. We define the odd function $\mathcal{M}_e(s)$ using the calculated $sM(s)$ and use a linear interpolation, formulated as $\mathcal{G}(s) = s\mathcal{M}_e(s_{min})/s_{min}$, as a first guess for the missing data at $s <$1.6 Å$^{-1}$. The support constraint is applied using a band-pass filter that selects the signal within the range $r_1 < r < r_2$ , given by

$$\mathcal{H}(\mathrm{r}) = e^{-\left(\frac{r-r_c}{w}\right)^{2\mathcal{N}}}, \quad (16)$$

where $r_c = (r_1 + r_2)/2$ and $w = (r_2 - r_1)/2$, and $\mathcal{N}$ is a positive integer. The sharpness of the filter is determined by $\mathcal{N}$ . This filter is preferable to a rectangular window because it avoids generating discontinuities at the edges. The parameters are set based on prior knowledge of the size of the molecule, with $r_1$ chosen to be smaller than the minimum interatomic distance, and $r_2$ larger than the longest interatomic distance.

We tested the performance of the algorithm using two cases: a static diffraction pattern of iodobenzene ($C_6H_5I$) and a difference-signal comparable to those used in GUED. For the difference signal we assumed photodissociation leading to $C_6H_5$ +I. The numerical calculation of $sM(s)$ or $\Delta sM(s)$ is based on eqn (1-4) and (6-7). The molecular structures were obtained from an online database [53]. The Matlab code for the iterative algorithm can be found in the Supplemental Material [54].

### *A. Static diffraction*

A model of the iodobenzene molecule is shown as an inset in Figure 2(a), in which the carbon, iodine and hydrogen atoms are represented by dark grey, purple and light gray colors, respectively. The minimum and maximum interatomic distances in iodobenzene are 1.090 Å and 5.953 Å. We defined the constraint function $\mathcal{H}(\mathrm{r})$ with $r_1$= 0.68 Å, $r_2$ = 6.20 Å and $\mathcal{N}$ = 15. The damping constant is $d = 0.01$ Å$^2$.

The input and restored signals are shown in Figure 2. In the 1st iteration, we set $\widetilde{\mathcal{M}}_1(s) = \mathcal{M}_e(s)$ and fill the missing data from 0 Å$^{-1}$ to 1.6 Å$^{-1}$ using a linear function. The $\mathcal{P}_1(r)$ and $\mathcal{P}(r)$, which are the FST of $\widetilde{\mathcal{M}}_1(s)$ and $\mathcal{M}(s)$, respectively, are shown in Figure 2(b). The missing data introduces significant artifacts in real space representation $\mathcal{P}_1(r)$. The band-pass filter used to apply the support constraint, $\mathcal{H}(\mathrm{r})$, is also shown in Figure 2(b). Figure 2(c)-(d) show the retrieved $\widetilde{\mathcal{M}}_{120}(s)$ and $\mathcal{P}_{120}(r)$ after 120 iterations, which match the true signals very well. We also tested the iterative algorithm with multiple different filling methods, such as cubic interpolation, next neighbor interpolation, and nearest neighbor interpolation, and did not see difference in the retrieved results. This indicates the retrieval is not dependent on the initial guess.

The retrieval error for the simulated data is tracked using the sum of the square of the residuals in the region of missing data ($s < s_{min}$) between $\widetilde{\mathcal{M}}_n(s)$ and $\mathcal{M}(s)$:

$$\mathcal{R}_n = \int_0^{s_{min}} \left[\widetilde{\mathcal{M}}_n(s) - \mathcal{M}(s)\right]^2 ds / s_{min}. \quad (17)$$

Figure 3 shows both versions of the retrieval error (the left ordinate is for $\mathcal{S}_n$, and right ordinate for $\mathcal{R}_n$). Note that $\mathcal{R}_n$ is only accessible for simulated data, while the function $\mathcal{S}_n$ can be used for experimental data where the true signal is unknown. The inset shows that $\mathcal{R}_n$ is converging more slowly than $\mathcal{S}_n$, with both $\mathcal{S}_n$and $\mathcal{R}_n$ approaching a small value that is close to zero after 60 iterations. We also tested the iterative algorithm with the simulated diffraction signal of a smaller molecule, trifluoroiodomethane ($CF_3I$), in APPENDIX A. In this case we observed that both $\mathcal{S}_n$ and $\mathcal{R}_n$ converge at the same rate.

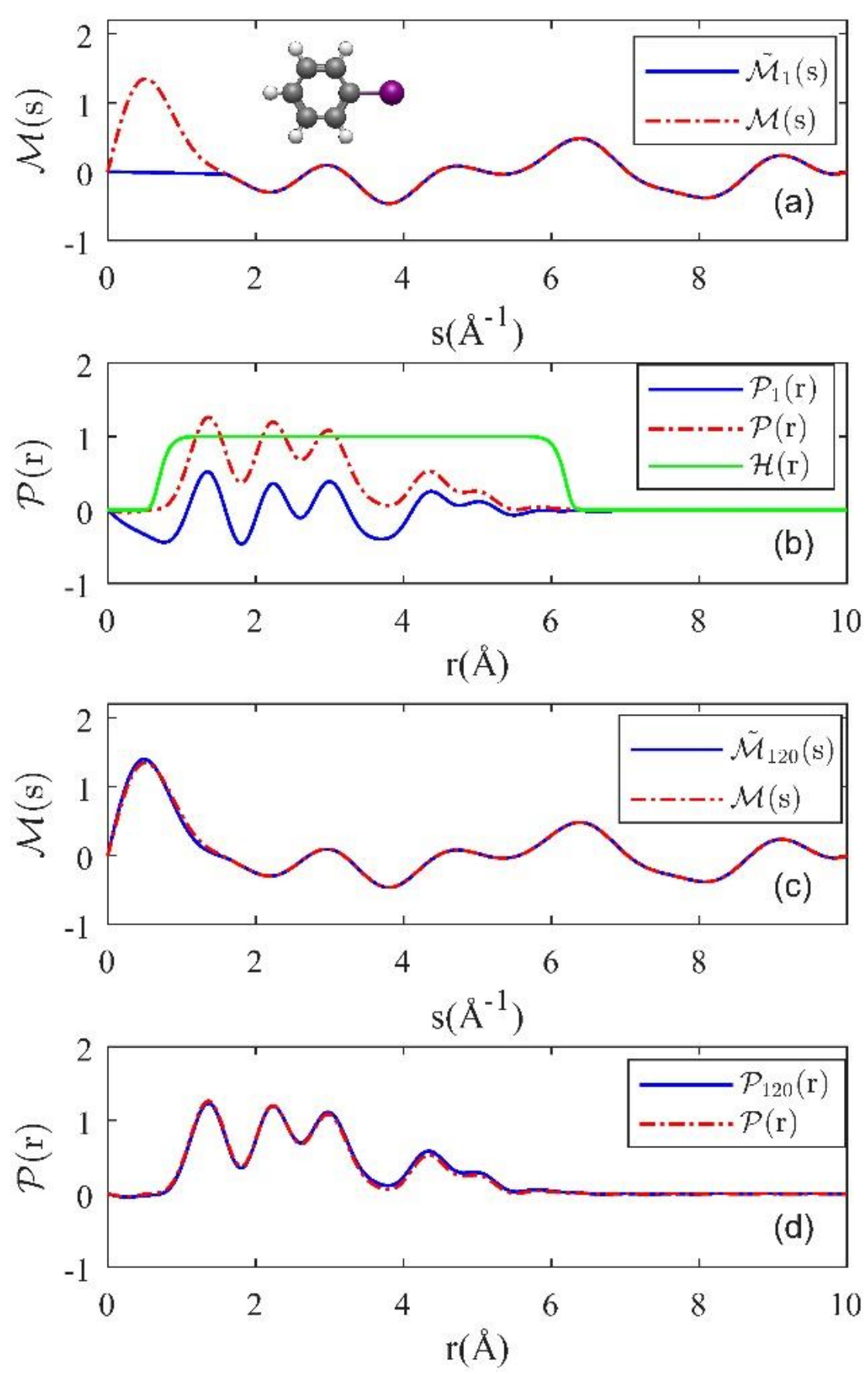

Figure 2. Retrieval of the diffraction signal of iodobenzene. (a) $\widetilde{\mathcal{M}}_1(s)$ is the first guess where the missing signal was filled by a linear interpolation (solid blue line) and $\mathcal{M}(s)$ is the true signal (dashed red line). The inset shows a model of the iodobenzene molecular structure, where the carbon atoms are shown in dark grey, the iodine atom in purple, and the hydrogen atoms in light grey. (b) $\mathcal{P}_1(r)$ (solid blue line) and $\mathcal{P}(r)$ (dashed red line) produced by FST. $\mathcal{H}$(r) (solid green line) is the band-pass filter used to apply the support constraint. (c) The restored signal after 120 iterations, $\widetilde{\mathcal{M}}_{120}(s)$ (solid blue line) and the true signal $\mathcal{M}(s)$ (dashed red line). (d) The corresponding $\mathcal{P}_{120}(r)$ (solid blue line) and the $\mathcal{P}(r)$ (dashed red line).

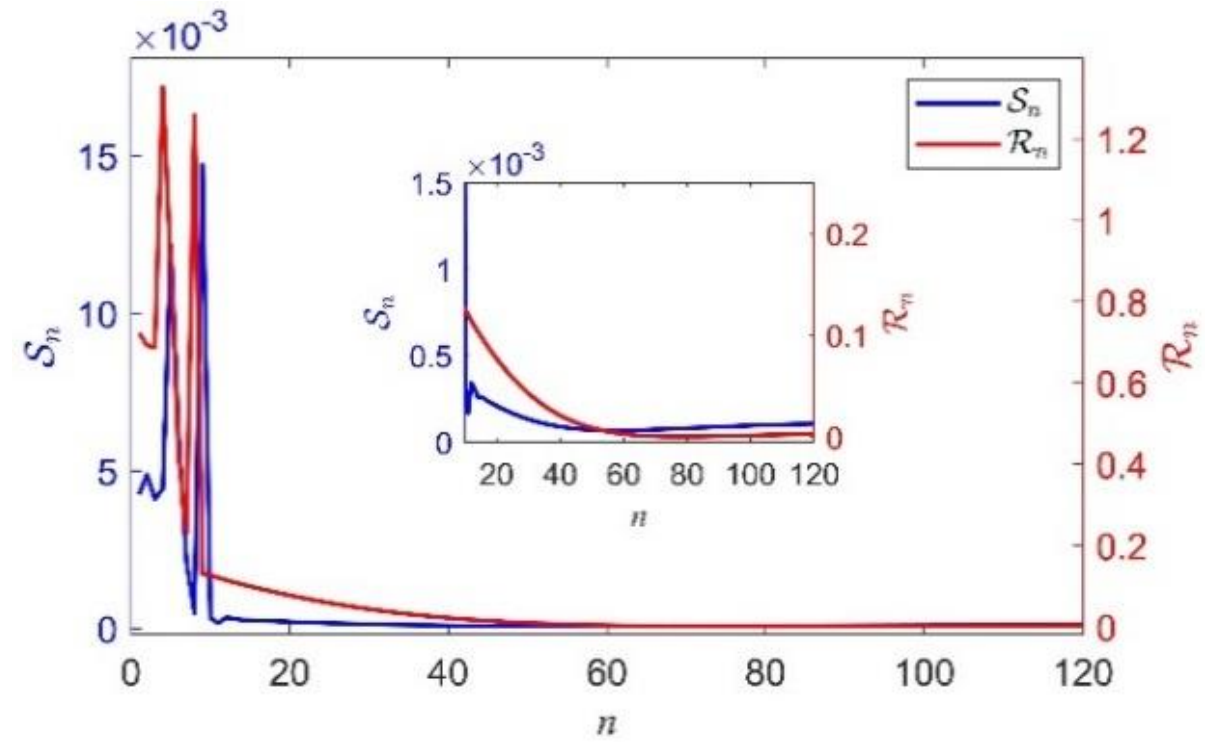

Figure 3. The error functions $\mathcal{S}_n$ and $\mathcal{R}_n$ as a function of iteration number *n*.

### *B. Dissociated iodobenzene*

We now consider the case of time-resolved measurements in which a signal after laser excitation is compared with a signal before the laser excitation. In this example we consider the dissociation of the iodine atom in iodobenzene, $C_6H_5I \rightarrow C_6H_5 + I$, and for simplicity assume that the structure of the phenyl ring remains unchanged. We calculated the diffraction-difference intensity, formulated as $\Delta I_{\mathrm{M}}(s) = I_{\mathrm{C_6H_5}}(s) + I_{\mathrm{I}}(s) - I_{\mathrm{C_6H_5I}}(s)$, where $I_{\mathrm{C_6H_5}}(s)$, $I_{\mathrm{I}}(s)$, $I_{\mathrm{C_6H_5I}}(s)$ are the total scattering intensities of the phenyl ring, an iodine atom and iodobenzene calculated using eqn (1)-(3). After the dissociation the scattering from the iodine atom no longer interferes with the scattering from the other atoms. The $\Delta sM(s)$ is calculated using eqn (7), and the $\mathcal{M}(s)$ is defined with eqn (8b). We used eqn (11) to define $\mathcal{M}_e(s)$ with $\Delta sM(s)$ from 1.6 Å$^{-1}$ to 7.5 Å$^{-1}$.

The input and true signals are shown in Figure 4(a-b). The function $\mathcal{H}$(r) was set with parameters $r_1$ = 1.15 Å, $r_2$ = 6.20 Å and $\mathcal{N} = 12$ . The damping constant is $d = 0.01$ Å$^2$. Here, it is important to point out that because these are difference-signals, positive values in $\mathcal{P}(r)$ indicate new distances that were created after the reaction, while negative values result from distances that are missing after the reaction. The artifacts introduced by the missing data can shift and offset the real space signal making it difficult to distinguish whether the peaks truly represent new distances, which makes the data interpretation challenging. In this case, because the reaction is a photodissociation and the structure of the phenyl ring is assumed to remain unchanged, there are no positive signals in $\mathcal{P}(r)$, as opposed to the $\mathcal{P}_1(r)$ which has both positive and negative regions. Figure 4(c-d) shows the retrieved signals in momentum space and real space, respectively, after 150 iterations. The difference between $\widetilde{\mathcal{M}}_{150}(s)$ and $\mathcal{M}(s)$ is significantly reduced, and the restored data from 0 Å$^{-1}$ to 1.6 Å$^{-1}$ is in good agreement with the original signal $\mathcal{M}(s < 1.6\ \text{Å}^{-1})$. The retrieved real space signal $\mathcal{P}_{150}(r)$ is in excellent agreement with the true signal $\mathcal{P}(r)$.

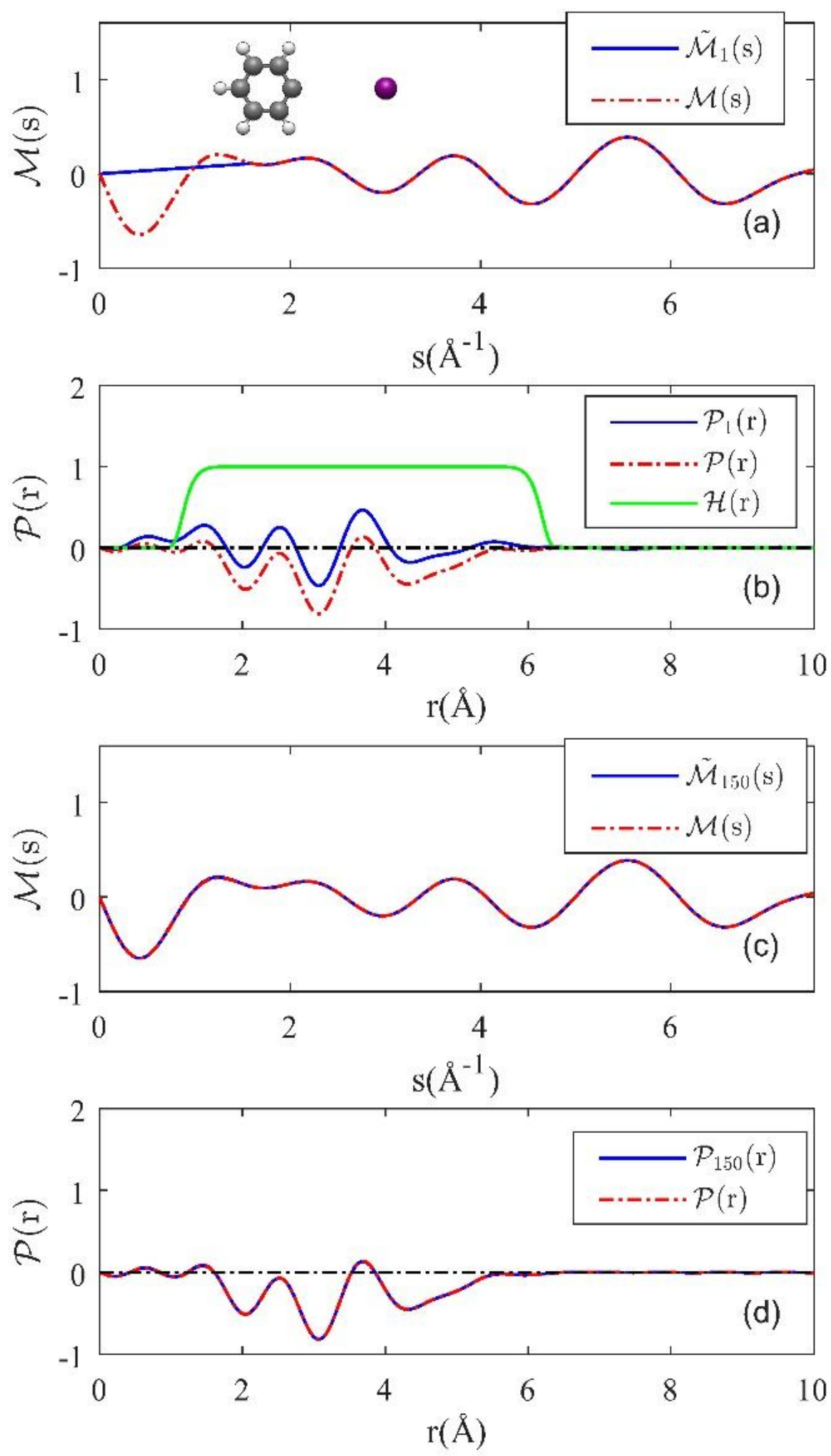

Figure 4. Retrieval from the difference signal of dissociated iodobenzene. (a) The first guess $\widetilde{\mathcal{M}}_1(s)$ (solid blue line) and the true signal $\mathcal{M}(s)$ (dashed red line). The inset shows a model of the structure after dissociation. (b) $\mathcal{P}_1(r)$ (solid blue line) and $\mathcal{P}(r)$ (dashed red line) generated by FST of the functions in panel (a). The band-pass filter $\mathcal{H}$(r) is shown by the solid green line. The zero level is shown by the dashed black line. (c) The restored signal after 150 iterations, $\widetilde{\mathcal{M}}_{150}(s)$, is shown by the solid blue line. The dashed red line shows the true signal $\mathcal{M}(s)$. (d) The FST of $\widetilde{\mathcal{M}}_{150}(s)$ produces $\mathcal{P}_{150}(r)$, shown in the blue line. The dashed red line shows the true real-space signal $\mathcal{P}(r)$.

The behavior of $\mathcal{S}_n$ and $\mathcal{R}_n$ is shown in Figure 5. Note that the left y axis applies to $\mathcal{S}_n$, and right y axis to $\mathcal{R}_n$. Both $\mathcal{S}_n$and $\mathcal{R}_n$approach zero after 50 iterations. We examine the convergence behavior in more detail in the Discussion section.

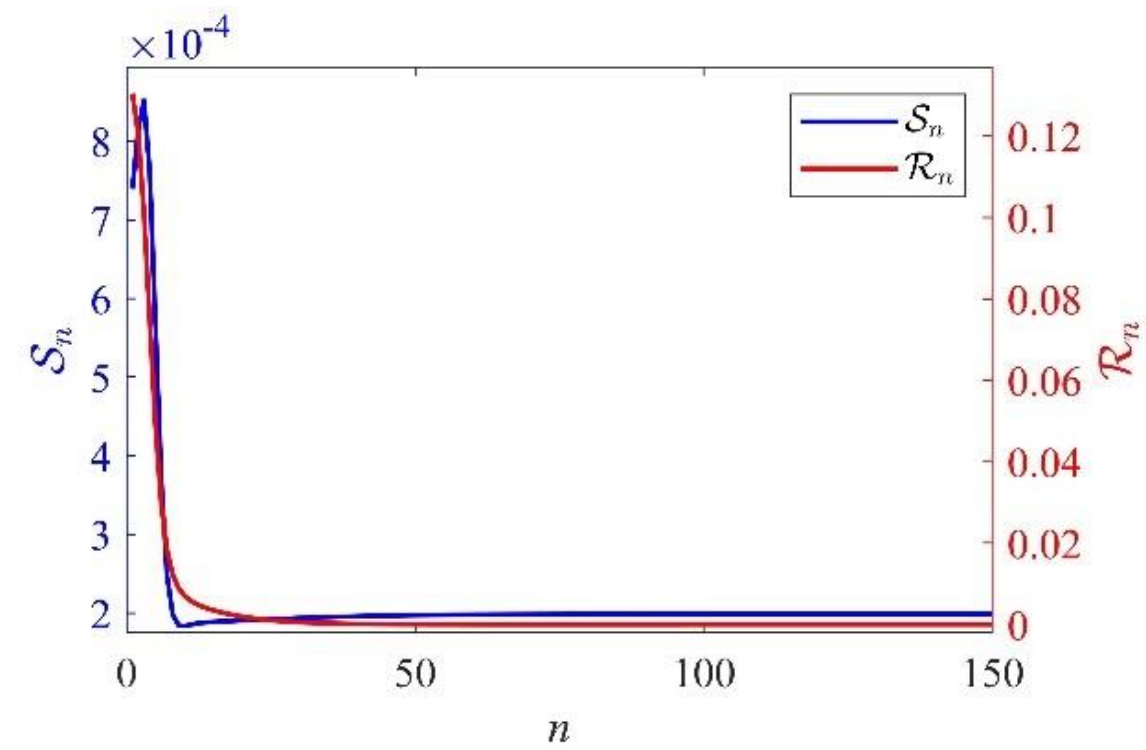

Figure 5. Convergence behavior of $\mathcal{S}_n$ and $\mathcal{R}_n$ in retrieving the difference diffraction signal of dissociated iodobenzene. *n* is the iteration number.

## IV. APPLICATION TO EXPERIMENTAL DATA

In this section, we apply the iterative algorithm to experimental data on the dissociation of iodobenzene to test the performance in the presence of noise. Here we assume that the structure of the ground state is known but the structure of the dissociated molecule is not, which is typically the case in GUED and UXRD experiments. The experiment was conducted using a table-top keV-UED instrument, which has been described in detail previously [5, 52, 55]. A femtosecond 266 nm ultraviolet (UV) laser pulse was used to photo-dissociate the molecules, and diffraction patterns were recorded both before and after the reaction to generate the diffraction-difference signal. The time delay between the laser excitation and the arrival of the electron pulse was >1 ps, after the reaction is completed. The 2-D diffraction-difference patterns were azimuthally averaged to obtain the one-dimensional $\Delta I(s)$ and $\Delta I/I_0$ , where $I_0(s)$ is a reference signal corresponding to unexcited molecules. Residual background in the difference signal was removed by fitting a third order polynomial and then subtracting it from the $\Delta I/I_0$ , to obtain the experimental $\Delta sM$ using the equation $\Delta sM =$ $\left(\frac{\Delta I}{I_0}\right)_{exp} \times \left(\frac{sI_0}{I_A}\right)_{th}$, where $\left(\frac{sI_0}{I_A}\right)_{th}$ is a calculated signal based on the theoretical ground state structure.

The minimum C-C distance of iodobenzene in the ground state is 1.395 Å, and minimum C-H distance is 1.087 Å. The longest distance is the H-I distance for the furthest H atom, corresponding to 5.953 Å. The parameters for the bandpass function $\mathcal{H}(\mathrm{r})$ are $r_1$ = 1.25 Å, $r_2$ = 6.10 Å and $\mathcal{N} = 12$. We select $r_1$ to be smaller than the C-C distance of 1.395 Å, which could be affected by vibrations in the ring. We do not expect significant changes in the C-H distances after photo-dissociation, so we do not set $r_1$ shorter than the C-H distance. Using a tighter constraint speeds up the retrieval and produces more accurate results.

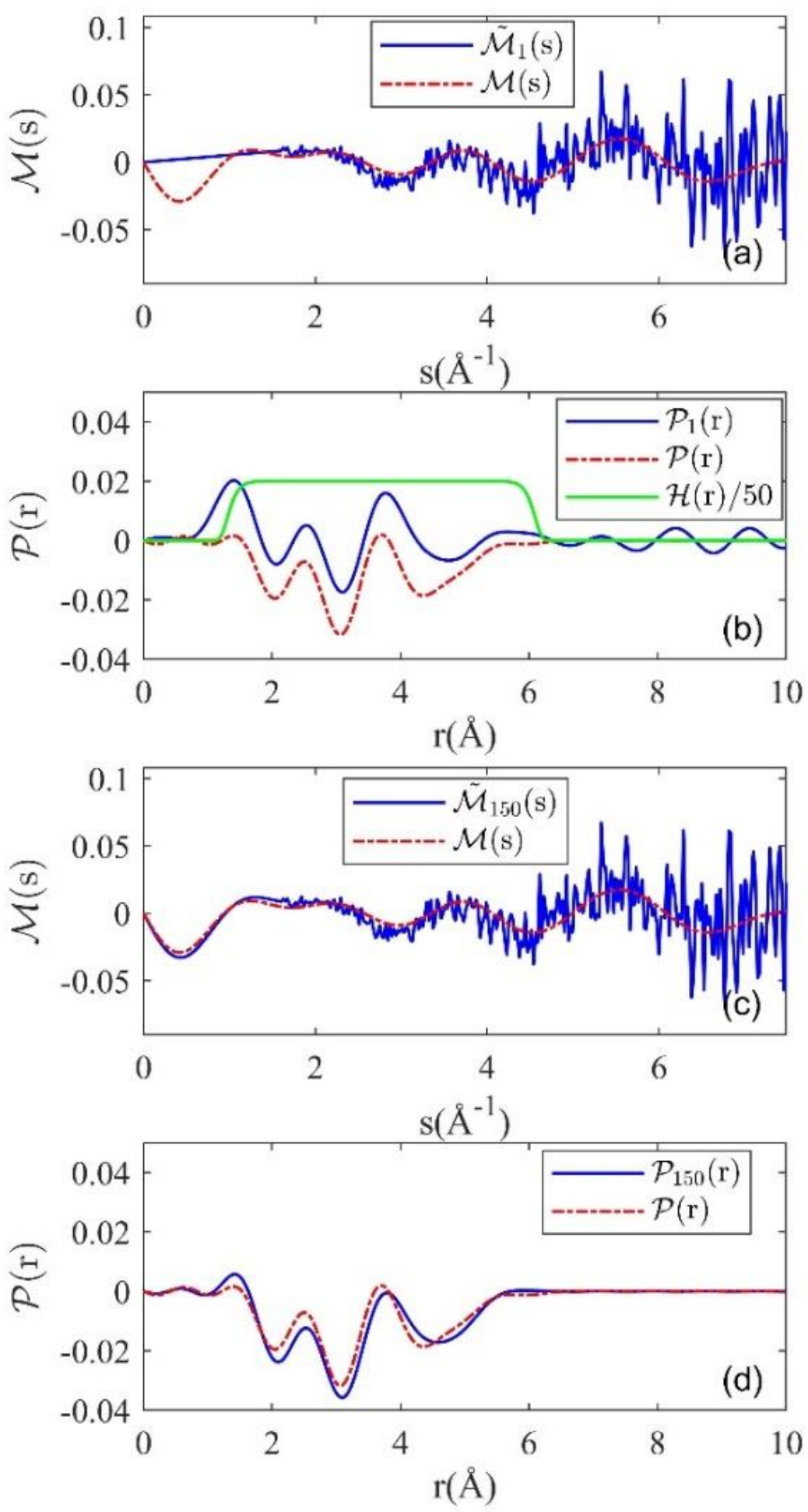

Figure 6. Retrieval from the difference signal in dissociated iodobenzene (a) The first guess $\widetilde{\mathcal{M}}_1(s)$ (solid blue line) and the theoretical signal $\mathcal{M}(s)$ (dashed red line). (b) $\mathcal{P}_1(r)$ (solid blue line) and $\mathcal{P}(r)$ (dashed red line) generated by FST of the functions in panel (a). The band-pass filter $\mathcal{H}(\mathrm{r})$ is shown by the solid green line. (c) The restored signal after 150 iterations, $\widetilde{\mathcal{M}}_{150}(s)$, is shown by the solid blue line. The dashed red line shows the theoretical signal $\mathcal{M}(s)$. (d) The FST of $\widetilde{\mathcal{M}}_{150}(s)$ produces $\mathcal{P}_{150}(r)$, shown by the solid blue line. The dashed red line shows the theoretical real-space signal $\mathcal{P}(r)$.

The input and retrieved signals are shown in Figure 6, along with theoretical signals for comparison. The available experimental signal for the first guess $\widetilde{\mathcal{M}}_1(s)$ is from 1.6 Å$^{-1}$ to 7.5 Å$^{-1}$ and the missing data is filled using a linear function $\mathcal{G}(s) = s\mathcal{M}_e(s_{min})/s_{min}$. The noise level is higher at high values of $s$ because the scattering amplitude decreases rapidly with increasing $s$, and the difference signal is only a small fraction of the total signal. For the theory comparison, we have the same model that was used in section III.B to approximate the dissociation signal, where we assume the structure of the phenyl ring remains unchanged. This ignores vibrational excitation of the ring but serves as a reasonable guide because the strongest contribution to the difference signal comes from removal of the iodine atom, which has a much larger scattering cross section than carbon and hydrogen. The $\mathcal{M}(s)$ in Figure 6(a) was produced using this model. The calculated signal $\mathcal{M}(s)$ is shown only for reference and is not used to retrieve the missing region of the experimental signal. The corresponding real space signals $\mathcal{P}_1(r)$ and $\mathcal{P}(r)$, are shown in Figure 6(b). The bandpass function $\mathcal{H}(\mathrm{r})$ is shown by the green line in Figure 6(b). The damping constant is $d = 0.02$ Å$^2$. The retrieved $\widetilde{\mathcal{M}}_{150}(s)$ and $\mathcal{P}_{150}(r)$ are shown in Figure 6(c,d), respectively. The retrieved $\mathcal{P}_{150}(r)$ is in good agreement with the calculated $\mathcal{P}(r)$. The difference between the two could be due to vibrational motion of the ring or to residual background in the experimental signal. Figure 7 shows $\mathcal{S}_n$ as a function of iteration number ($n$), which approaches the minimum value after 50 iterations

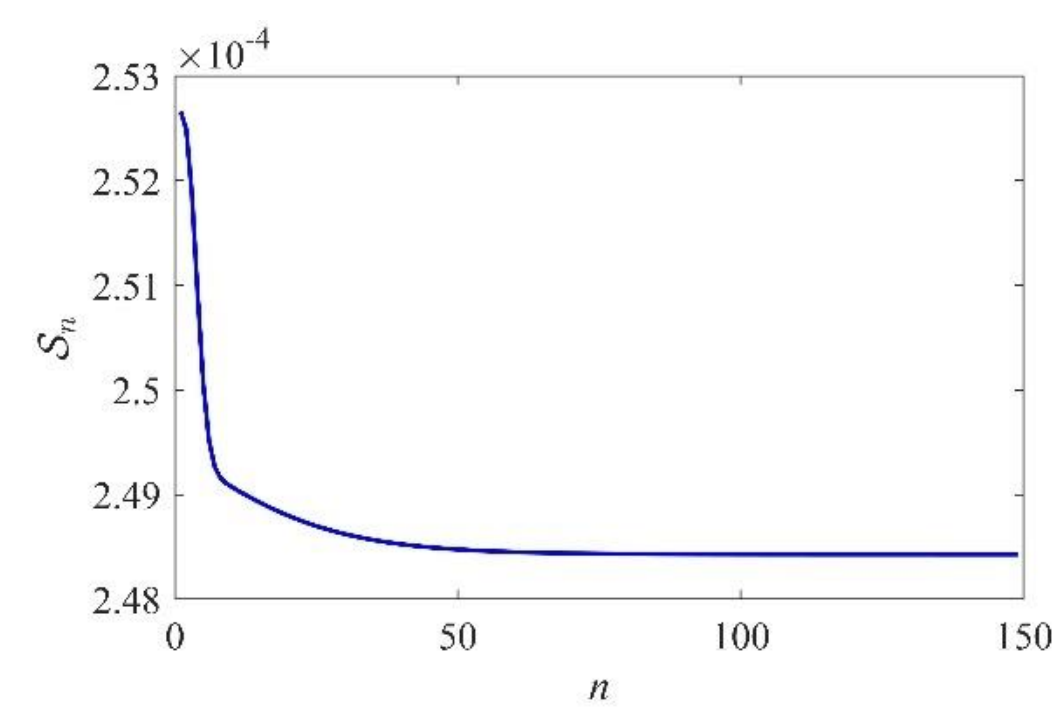

Figure 7. The function $\mathcal{S}_n$ computed in retrieving the photo-dissociation signal of iodobenzene. The iteration number is denoted as $n$.

## V. DISCUSSION

As shown in Figures 3 and 5, $\mathcal{R}_n$ converges slower than $\mathcal{S}_n$. The reason for this is that the distribution of the PDF signal is broad, meaning that a large component of the artifact overlaps with the true PDF and is not reduced by the constraint. We have tested also the convergence with a smaller molecule, trifluoroiodomethane, in which case $\mathcal{S}_n$ and $\mathcal{R}_n$ converge faster and at the same rate (see APPENDIX A for details). This convergence behavior indicates that the iterative algorithm can be expected to perform better for more compact molecules. Most molecules so far studied by GUED or UXRD are comparable in size or smaller than iodobenzene, so the algorithm is broadly applicable to these molecules.

Figure 3 also shows oscillations in both $\mathcal{S}_n$ and $\mathcal{R}_n$ over the initial iterations before settling down after about 20 iterations. For this reason, it is better to continue running the algorithm even after it reaches a small value to ensure that it has converged. This also indicates that the algorithm could find a "wrong" solution. That is, one that matches the data over the measured range but produces unphysical results over the missing range. We have observed that these solutions are very different from the correct signal and could be discarded based on producing unphysical results. This problem can also be resolved by making small changes in $s_{min}$ and checking if the solution changes. We have also tested that the algorithm works for a more limited *s* range, such as 1.0 Å$^{-1}$ to 4.0 Å$^{-1}$, typical of UXRD measurements [13, 56, 57] (see APPENDIX A and APPENDIX B).

It is important to set the limits of the filter function $\mathcal{H}(\mathrm{r})$ correctly. If the filter range is too small, it will cut some of the true signal and introduce artifacts, and if it is too large the retrieval will be less accurate. The recommended procedure is to start with a broad filter that is larger than the true signal in real space, and then iteratively reduce the filter size while monitoring the error function $\mathcal{S}_n$. As the filter size is decreased the error will initially decrease and then start to increase as it starts to cut some of the true signal. We have noticed that there is a range where the error is minimized and the retrieval result is independent of filter size, corresponding to the optimal setting of the filter.

The range of the filter function $\mathcal{H}(\mathrm{r})$ can also be estimated from the real-space signal $\mathcal{P}_1(\mathrm{r})$ calculated using the momentum-transfer signal with a guess for the missing low-*s* signal. This can be applied for cases where there are time dependent distances that are not known *a prior*, for example in a dissociation reaction. In this case, the function $\mathcal{P}_1(\mathrm{r})$ will show the approximate location of the dissociated wavepacket, which is sufficient to generate the required constraint function. In APPENDIX C, we show an example of retrieving the signal of $CF_3I$ with CI distance increased by 5Å with the filter parameters estimated by the real space signal $\mathcal{P}_1(\mathrm{r})$. In this case the filter would consist of two parts, one of which will be time-dependent to capture the increasing distance between the fragments. Alternatively, if the structure of the reaction products is completely unknown, the filter size can be adjusted iteratively starting from a large size. This "shrinkwrap" procedure has been demonstrated previously [43] and does not require any prior knowledge. A situation in which the retrieval might not work properly is if the reaction products include a broad range of new distances, for example a slow dissociation process in which the dissociation wavepacket is spatially very broad.

## VI. CONCLUSION

In conclusion, we report an iterative retrieval algorithm to restore missing data in time-resolved diffraction experiments. An important advantage of this algorithm is that it is simple to implement and requires only a minimal amount of prior knowledge. The missing data in the low momentum transfer region is essential for accurate interpretation of the real-space signal. Our algorithm requires only *a-priori* knowledge of the size of the molecule, *i.e.* the shortest and longest interatomic distances, which are known in most cases. For the case of time resolved experiments, a tighter constraint could be used based on the shortest and longest interatomic distances that change after the reaction, unless the maximum distance is increased by the reaction. The algorithm retrieves the PDF or ΔPDF of the products and thus can also be applied to reactions with multiple channels. Additionally, this algorithm could be applied to separate the electronic and nuclear contributions to GUED signals. The electronic contribution is limited to the low *s* region, so by removing this region of the data and then applying the retrieval algorithm, the missing data from nuclear motion alone would be retrieved, from which the electronic contribution can be isolated by comparing to the total signal. Here we have validated the method with both simulated and experimental GUED data of iodobenzene.

## ACKNOWLEDGMENTS

This work was supported by the US Department of Energy Office of Science, Basic Energy Sciences under award no. DE-SC0014170.

## DATA AVAILABILITY

The data that support the findings of this article are openly available in [58].

## APPENDIX A: SIMULATED X-RAY DIFFRACTION OF CF3I

In this section, we tested the iterative algorithm with the simulated X-ray scattering signal of a smaller molecule, $CF_3I$. The molecular structures were obtained from [59]. The total elastic X-ray scattering $I_{total}(s)$, atomic scattering $I_A(s)$, and molecular scattering $I_M(s)$ have the same formulas as eqn (1), (2) and (3) for GUED, with the atomic scattering amplitudes $f_j(s)$ replaced by X-ray elastic scattering amplitudes. We applied the iterative algorithm to the simulated X-ray diffraction signal assuming an X-ray photon energy of 15 keV. The useful data in typical UXRD experiments is usually from 0.5 Å$^{-1}$ to 4 Å$^{-1}$ [13, 56, 57]. To demonstrate the capability of the algorithm, the available momentum transfers of the simulated signal ranges from 1 Å$^{-1}$ to 4 Å$^{-1}$, and we aim to retrieve the signal between 0 Å$^{-1}$ and 1 Å$^{-1}$. The procedures are the same as those described in section III in the main text.

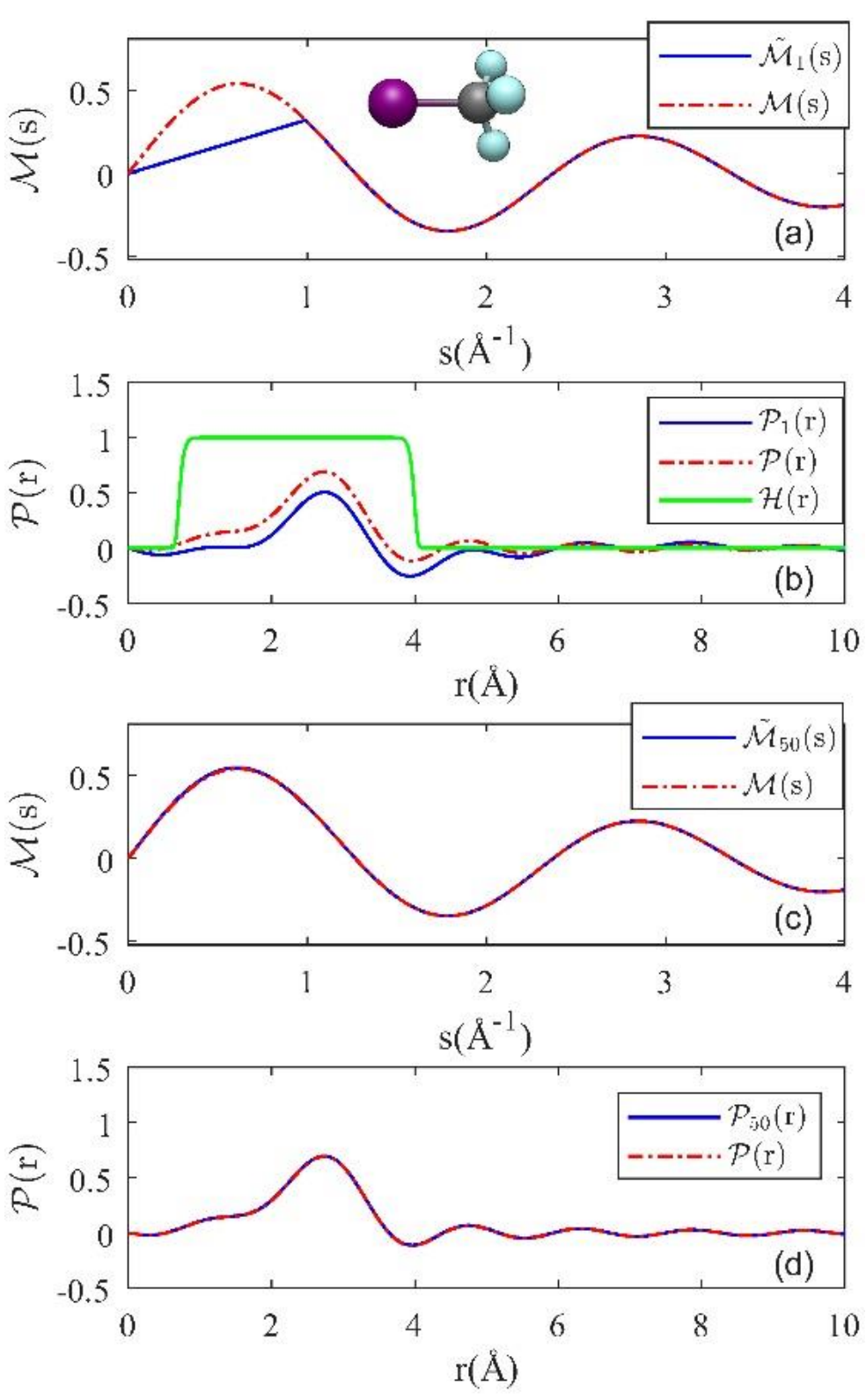

Figure 8. Retrieval from simulated X-ray diffraction data of $CF_3I$ molecules in the ground state with *s*-range from $1.0\text{Å}^{-1}$ to $4.0\text{Å}^{-1}$. (a) The first guess $\widetilde{\mathcal{M}}_1(s)$ (solid blue line) and the true signal $\mathcal{M}(s)$ (dashed red line). The inset shows a model of the $CF_3I$ structure, where the iodine atom is purple, the carbon atom is grey, and the fluorine atoms are green. (b) $\mathcal{P}_1(r)$ (solid blue line) and $\mathcal{P}(r)$ (dashed red line) generated by FST of the functions in panel (a). The band-pass filter $\mathcal{H}(\mathrm{r})$ is shown by the solid green line. (c) The restored signal after 50 iterations, $\widetilde{\mathcal{M}}_{50}(s)$, is shown by the solid blue line. The dashed red line shows the true signal $\mathcal{M}(s)$. (d) The FST of $\widetilde{\mathcal{M}}_{50}(s)$ produces $\mathcal{P}_{50}(r)$, shown in the blue line. The dashed red line shows the true real-space signal $\mathcal{P}(r)$.

The input and real signals in both momentum transfer and real spaces are shown in Figure 8(a-b). The inset of Figure 8 (a) shows a model of the $CF_3I$ molecular structure, where the iodine atom is purple, the carbon atom is grey, and the fluorine atoms are green. The band-pass filter is shown by the green line in Figure 8(b). Figure 8(c)-(d) shows that a very good agreement between the retrieved signal and true signal is achieved after 50 iterations.

The convergence behavior of $\mathcal{S}_n$ and $\mathcal{R}_n$ (left ordinate is for $\mathcal{S}_n$, and right ordinate is for $\mathcal{R}_n$) are shown in Figure 9. Both $\mathcal{S}_n$and $\mathcal{R}_n$ approach quickly to zero at the same rate after only 5 iterations.

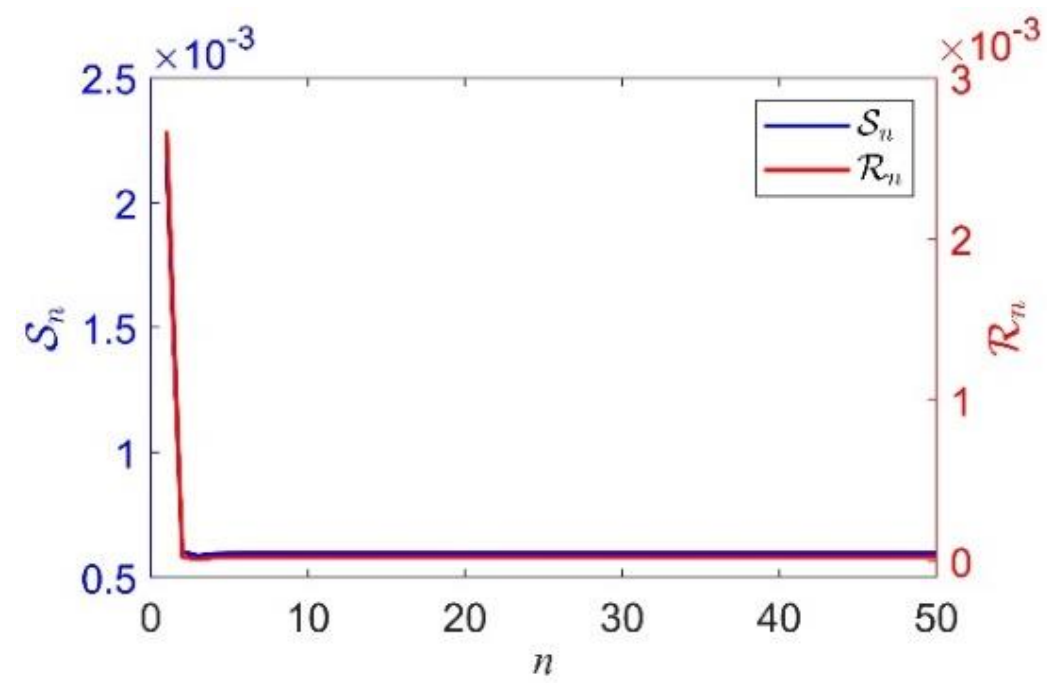

Figure 9. Convergence behavior of $\mathcal{S}_n$ and $\mathcal{R}_n$ in retrieving the diffraction signal of ground state $CF_3I$ molecules. The iteration number is denoted as *n*.

## APPENDIX B: SIGNAL OF IODOBENZENE WITH A SMALLER S-RANGE

In section III, we applied the iterative algorithm to successfully restore the signal from 0 $\text{Å}^{-1}$ to 1.6 $\text{Å}^{-1}$ using the simulated electron scattering signal of dissociated iodobenzene from 1.6 $\text{Å}^{-1}$ to 7.5 $\text{Å}^{-1}$, which corresponds to the *s*-range in typical GUED experiments. Here, we demonstrate that the algorithm can also be used to restore the low-*s* signal using the available signal with a smaller *s*-range.

The simulated signal of dissociated iodobenzene has been described in section III.B, and now we further limit the available signal to the range $1.6\text{Å}^{-1}$ to $4.0\text{Å}^{-1}$ and restore the signal from 0 $\text{Å}^{-1}$ to 1.6 $\text{Å}^{-1}$. The input and true signals are shown in Figure 10(a-b). The function $\mathcal{H}(\mathrm{r})$ was set with parameters $r_1$= 1.05 Å, $r_2$ = 6.20 Å and $\mathcal{N} = 12$ . The damping constant is $d =$ 0.012 Å$^2$. Figure 10(c-d) shows the retrieved signals in momentum space and real space, respectively, after 150 iterations. The difference between $\widetilde{\mathcal{M}}_{150}(s)$ and $\mathcal{M}(s)$ is significantly reduced, and the restored data from 0 Å$^{-1}$ to 1.6 Å$^{-1}$ is in good agreement with the true signal $\mathcal{M}(s < 1.6\text{ Å}^{-1})$. Although the signal is limited by a smaller region of momentum transfer, making the real space resolution significantly broader, as shown in Figure 10(b), the algorithm is still able to retrieve the correct signal in the low *s* region and reconstruct the ΔPDF accurately.

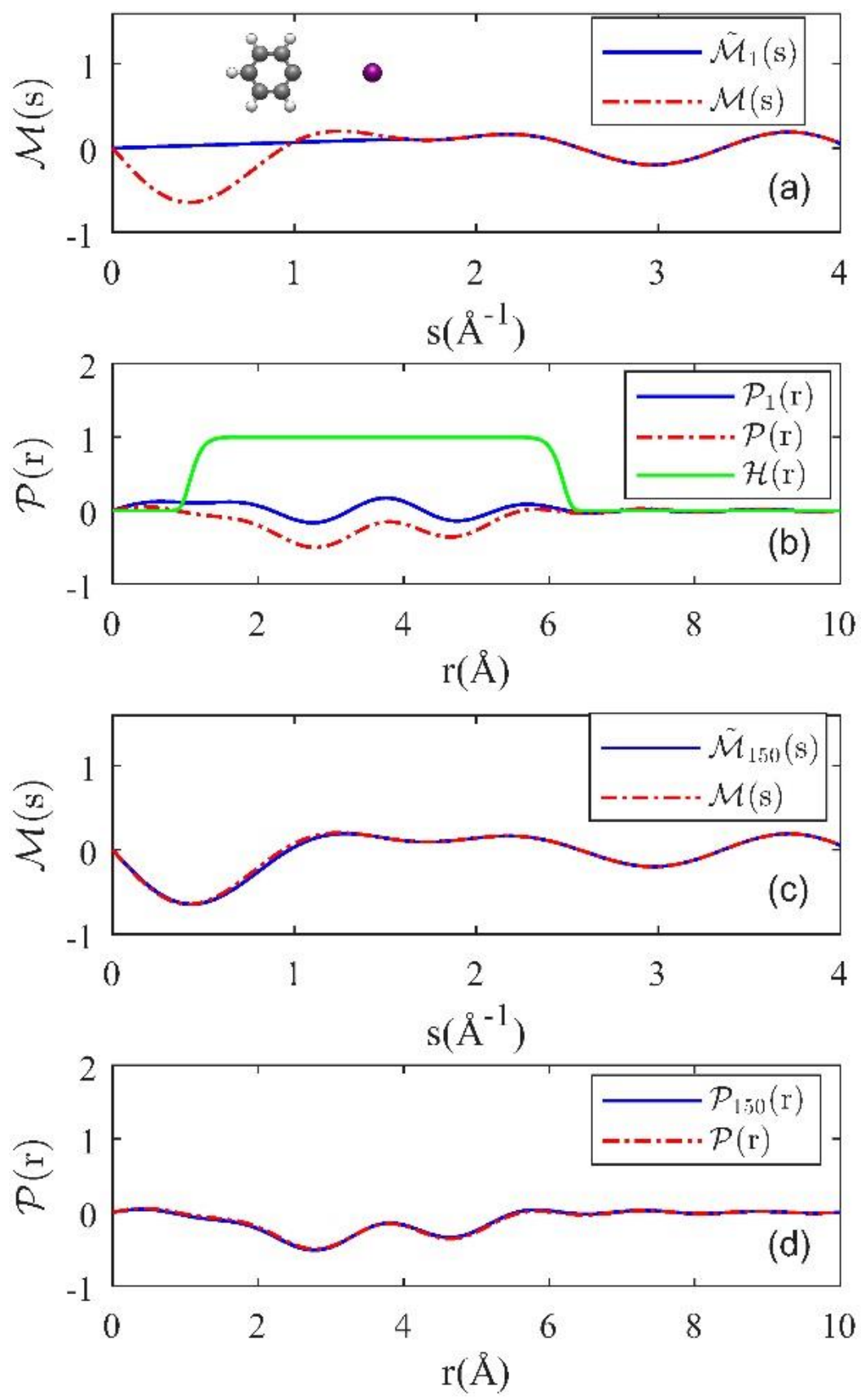

Figure 10. Retrieval from the difference signal in dissociated iodobenzene with *s*-range from 1.6Å$^{-1}$ to 4.0Å$^{-1}$. (a) The first guess $\widetilde{\mathcal{M}}_1(s)$ (solid blue line) and the true signal $\mathcal{M}(s)$ (dashed red line). The inset shows a model of the structure after dissociation. (b) $\mathcal{P}_1(r)$ (solid blue line) and $\mathcal{P}(r)$ (dashed red line) generated by FST of the functions in panel (a). The band-pass filter $\mathcal{H}(\mathrm{r})$ is shown by the solid green line. (c) The restored signal after 150 iterations, $\widetilde{\mathcal{M}}_{150}(s)$, is shown by the solid blue line. The dashed red line shows the true signal $\mathcal{M}(s)$. (d) The Fourier transform of $\widetilde{\mathcal{M}}_{150}(s)$ produces $\mathcal{P}_{150}(r)$, shown in the blue line. The dashed red line shows the true real-space signal $\mathcal{P}(r)$.

## APPENDIX C: SIGNAL OF $CF_3I$ DISSOCIATION

In this section, we consider the case of a dissociation reaction in which the distance between the fragments at any specific time is unknown. We show that the distance between the fragments can be estimated from $\mathcal{P}_1(r)$, without any prior knowledge. For this example, we simulated the electron diffraction signal of $CF_3I$ with CI distance increased by 5Å in the direction along the molecular axis, denoted as $\Delta\mathrm{CI} = 5$Å below, and for simplicity we assume the $CF_3$ structure remains unchanged. We calculated the diffraction-difference intensity $\Delta I_M(s) = I_M(s, \Delta\mathrm{CI} = 5\text{Å}) - I_M^0(s)$, where $I_M^0(s)$ is the molecular scattering of $CF_3I$ in the ground state.

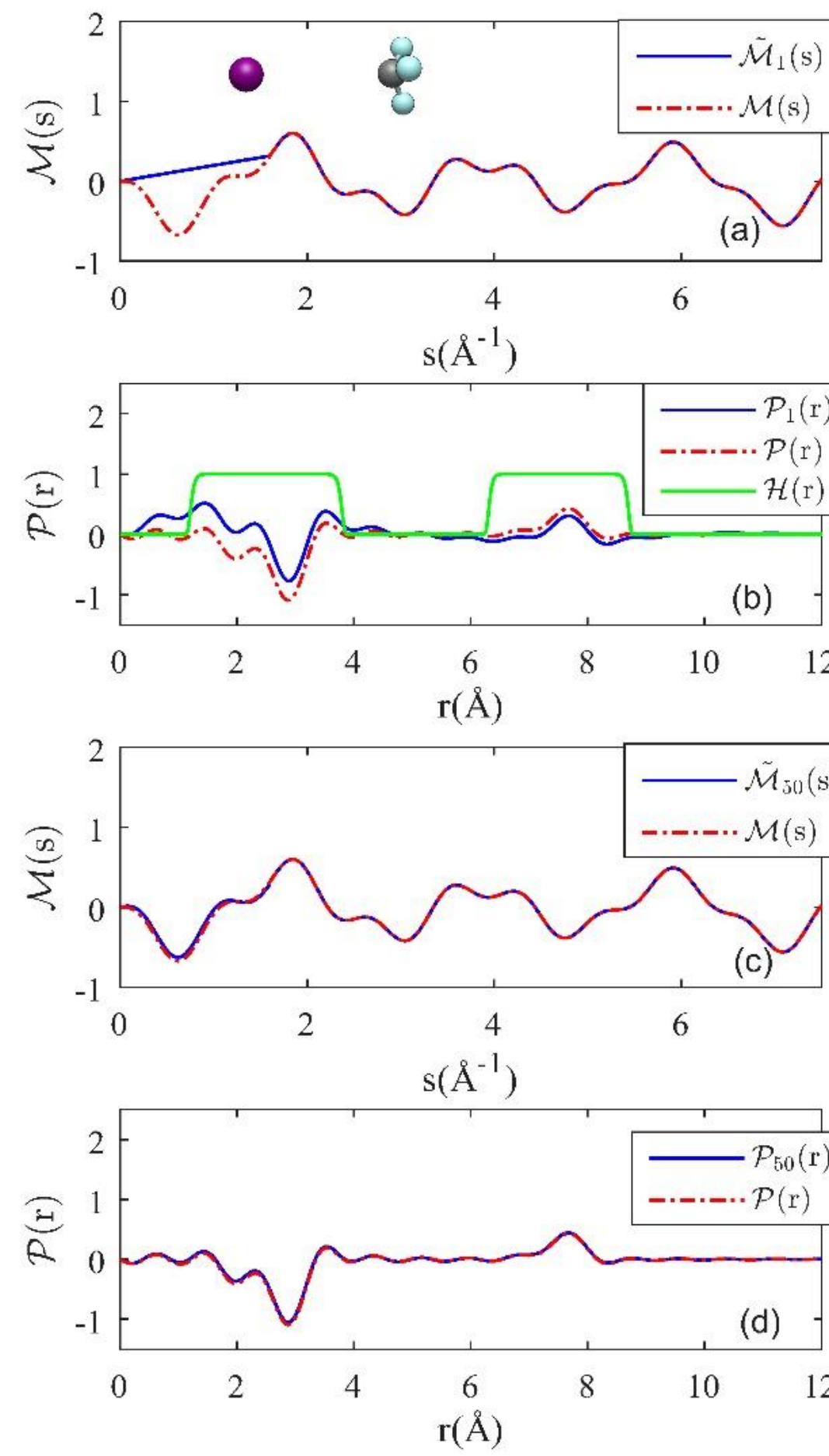

Figure 11. Retrieval from simulated diffraction data of $CF_3I$ molecules with the CI distance increased by 5Å. The signal from 1.6Å$^{-1}$ to 7.5Å$^{-1}$ is used to restore the missing signal in the range of 0Å$^{-1}$ to 1.6Å$^{-1}$. (a) The first guess $\widetilde{\mathcal{M}}_1(s)$ (solid blue line) and the true signal $\mathcal{M}(s)$ (dashed red line). The inset shows a model of the $CF_3I$ structure, where the iodine atom is purple, the carbon atom is grey, and the fluorine atoms are green.

(b) $\mathcal{P}_1(r)$ (solid blue line) and $\mathcal{P}(r)$ (dashed red line) generated by FST of the functions in panel (a). The band-pass filter $\mathcal{H}(\mathrm{r})$ is shown by the solid green line. (c) The restored signal after 50 iterations, $\widetilde{\mathcal{M}}_{50}(s)$, is shown by the solid blue line. The dashed red line shows the true signal $\mathcal{M}(s)$. (d) The FST of $\widetilde{\mathcal{M}}_{50}(s)$ produces $\mathcal{P}_{50}(r)$, shown in the blue line. The dashed red line shows the true real-space signal $\mathcal{P}(r)$.

The input and true signals are shown in Figure 11(a-b). In $\mathcal{P}_1(r)$, the two valleys at 2.14Å and 2.89Å are due to the loss of the two interatomic distance $r_{\mathrm{CI}}$ and $r_{\mathrm{FI}}$ of $CF_3I$ in the ground state. Due to the CI elongation by 5 Å, two new interatomic distances (CI and FI) are generated at 7Å and 7.7Å. Due to the limited spatial resolution and the stronger amplitude of the FI peak, a single peak is visible. This peak can be used to set a second part of the bandpass filter, given by $\mathcal{H}(\mathrm{r}) = \mathcal{H}_1(\mathrm{r}) + \mathcal{H}_2(\mathrm{r})$, where both $\mathcal{H}_1(\mathrm{r})$ are $\mathcal{H}_2(\mathrm{r})$ are given by eqn (16). For $\mathcal{H}_1(\mathrm{r})$, $r_1$= 1.2 Å, $r_2$ = 3.8 Å; and for $\mathcal{H}_1(\mathrm{r})$, $r_1$= 6.3 Å, $r_2$ = 8.7 Å. The band-pass filter $\mathcal{H}(\mathrm{r})$ is the green line, and $\mathcal{P}(r)$ is the dashed red line in figure 11(b). The other parameters are: $\mathcal{N} = 12$ and $d = 0.015$ Å$^2$. Figure 11(c-d) shows the retrieved signals in momentum space and real space, respectively, after 50 iterations. The difference between $\widetilde{\mathcal{M}}_{50}(s)$ and $\mathcal{M}(s)$ is significantly reduced, and the restored data from 0 Å$^{-1}$ to 1.6 Å$^{-1}$ is in good agreement with the true signal $\mathcal{M}(s < 1.6$ Å$^{-1})$. The $\mathcal{P}_{50}(r)$ and $\mathcal{P}(r)$ are in good agreement as well. Note that in this example represents the dissociation reaction at a specific time, and to apply this to time-resolved diffraction data the range of $\mathcal{H}_2(\mathrm{r})$ will be different for each time delay, while $\mathcal{H}_1(\mathrm{r})$ remains constant.